\documentclass[twocolumn,showpacs,preprintnumbers,amsmath,amssymb]{revtex4}

\usepackage[pdftex]{graphicx}

\usepackage{graphicx}
\usepackage{dcolumn}
\usepackage{bm}

\begin{document}


\title{A Universal Trimer in a Three-Component Fermi Gas} 

\author{A. N. Wenz}
\email{andre.wenz@mpi-hd.mpg.de}
\author{T. Lompe}
\author{T. B. Ottenstein}
\author{F. Serwane}
\author{G. Z\"urn}
\author{S. Jochim}
\altaffiliation[Also at:]{ Fakult\" at f\" ur Physik und Astronomie, \\
Ruprecht-Karls-Universit\"at Heidelberg, Germany}

%
\affiliation{%
Max-Planck-Institut f\" ur Kernphysik, Saupfercheckweg 1, \\
69117 Heidelberg, Germany
}

\date{1 October 2009}

\begin{abstract}

We show that the recently measured magnetic field dependence of three-body loss in a three-component mixture of ultracold $^6\rm{Li}$ atoms \cite{ottenstein, ohara} can be explained by the presence of a universal trimer state. Previous work suggested a universal trimer state as a probable explanation, yet failed to get good agreement between theory and experiment over the whole range of magnetic fields. For our description we adapt the theory of Braaten and Hammer \cite{braaten_review} for three identical bosons to the case of three distinguishable fermions by combining the three scattering lengths $a_{12},$ $a_{23}$ and $a_{13}$ between the three components to an effective interaction parameter $a_m$.  We show that taking into account a magnetic field variation in the lifetime of the trimer state is essential to obtain a complete understanding of the observed decay rates.
\end{abstract}

\maketitle

The fact that the long-range behavior of a system can become independent of the details of its short-range characteristics is a fundamental and powerful concept in physics, which is called universality. If a system has universal properties, these can be described by effective parameters without detailed knowledge of the short-range physics. 
Starting from this concept, Efimov \cite{efimov, efimov33b, efimov_radial_law} studied  three-particle interactions in a universal context and found that for diverging scattering length there is an infinite number of three-body bound states whose binding energies differ by a discrete scaling factor of $e^{- 2\pi /s_0}$, where $s_0$ is a numerical constant \cite{scaling_parameter}. For negative scattering lengths ($a<0$), these universal trimer states cross the three-atom continuum at critical values of $a$ spaced by $e^{ \pi /s_0} \approx 22.7$. 
For decades it remained ambiguous whether such universal three-body bound states actually exist and how they could be observed. 

This changed when ultracold gases provided a system in which the scattering length could be easily tuned over a wide range using magnetic Feshbach resonances.
This led to experimental evidence for the existence of Efimov states in an ultracold sample of $^{133}$Cs atoms by observing a resonantly enhanced three-body recombination rate which can be explained by a trimer state crossing the three-atom continuum \cite{innsbruck_efimov}.
Since then, universal few-body phenomena have been observed in several experiments with different bosonic systems \cite{innsbruck_atomdimer, lens_hetero, lens_homo}. Additionally, signatures for universal four-body bound states were observed as well \cite{lens_homo, greene_four-body, innsbruck_four-body, esry_four-body}.
Recently, three-body resonances were found in a system consisting of fermionic atoms in three different hyperfine states, indicating the presence of Efimov physics \cite{ottenstein, ohara}. Models describing these experimental data with a single Efimov-like trimer state crossing the continuum at two magnetic field values have shown good agreement for one resonance \cite{braaten_ourdata, naidon_ourdata, wetterich_ourdata}, yet it has not been possible to get full agreement over the whole magnetic field range. This left some doubt about the validity of the universal description of this system. 
In this Rapid Communication we argue that the reason for this discrepancy is not a breakdown of the universal theory.  Instead, we suggest that the assumption of a constant lifetime of the Efimov trimer over the whole magnetic field range made in those models is not valid in this case. Therefore we introduce a straightforward modification of the Braaten-Hammer model for three-body recombination in systems with large negative scattering lengths \cite{braaten_review} to include a magnetic field dependent lifetime of the trimer. This allows us to get full quantitative agreement with the experimental data for the whole magnetic field range, thereby showing that the observed loss can indeed be explained by the presence of a single universal trimer state.

 Three-body loss is quantitatively described by the three-body loss coefficient $K_3$ defined by
\begin{equation}
\dot{n}(\textit{\textbf{r}}) = - K_3 \: n^3 (\textit{\textbf{r}}),
\end{equation}
where $n(\textit{\textbf{r}})$ is the number density.
For a sample consisting of identical bosons with negative scattering length the three-body loss coefficient for recombination into a deeply bound dimer \cite{deep_dimer} and a free atom was derived in the zero-range and low-energy limit by Braaten and Hammer \cite{braaten_review} to be
\begin{equation}
K_{3,\rm{deep}} \left( a \right)=  \frac{c \, \sinh\left(2\, \eta_*\right)}{\sin^2 \left[s_0 \, \ln\left( a /  a_* \right)\right] + \sinh^2 \eta_*} \, \frac{\hbar \, a^4}{m},
\label{K_3_braaten}
\end{equation}
where $a$ is the two-body scattering length, c is a numerical constant and $s_0=1.00624$ is the scaling parameter \cite{low_energy}. All information about the three-body interaction potential and the decay of the trimer relevant for three-body recombination is described by the parameters $a_*$ and $\eta_*$. Every time the two-body scattering length fulfills $a= \left(e^{\pi/s_0}\right)^n \,a_*$ $\left(n=0,1,2,\dots\right)$ a universal three-body bound state crosses the continuum and leads to a resonant enhancement of three-body loss. The three-body parameter $a_*$ describes the effects of the short-range characteristics of the interaction potential on the long-range three-particle wave function and thus fixes the position of the loss resonances.
The width of the resonance is described by $\eta_*$ and is related to the lifetime of the Efimov state.  

\begin{figure} [h!]
\centering
	\includegraphics [width= 8.5cm] {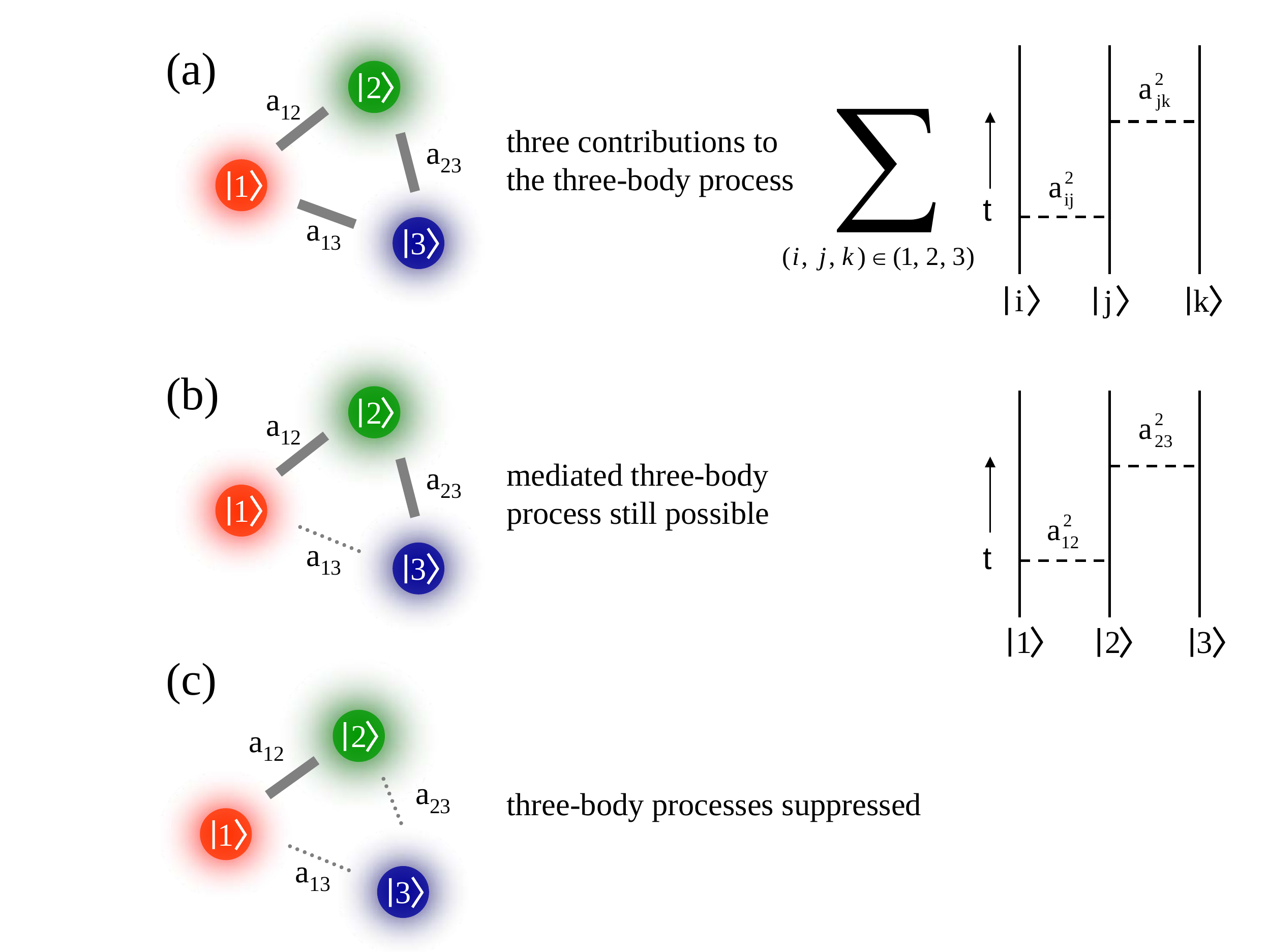}
	\caption{(Color online) A sketch of possible three-body processes in a system of three distinguishable fermions. In the case of three large scattering lengths, three different mediated processes have to be taken into account to describe the three-body interaction (see (a) and equation \ref{a_mean}). If one scattering length is significantly smaller than the other two, three-body processes are still possible but there is only one leading contribution (see (b)). If two scattering lengths are small compared to the third one (see (c)), three-body processes are suppressed compared to the situations described in (a) and (b).}
	\label{fig:mediated}
\end{figure}

\begin{figure} [htbp]
\centering
	\includegraphics [width= 8cm] {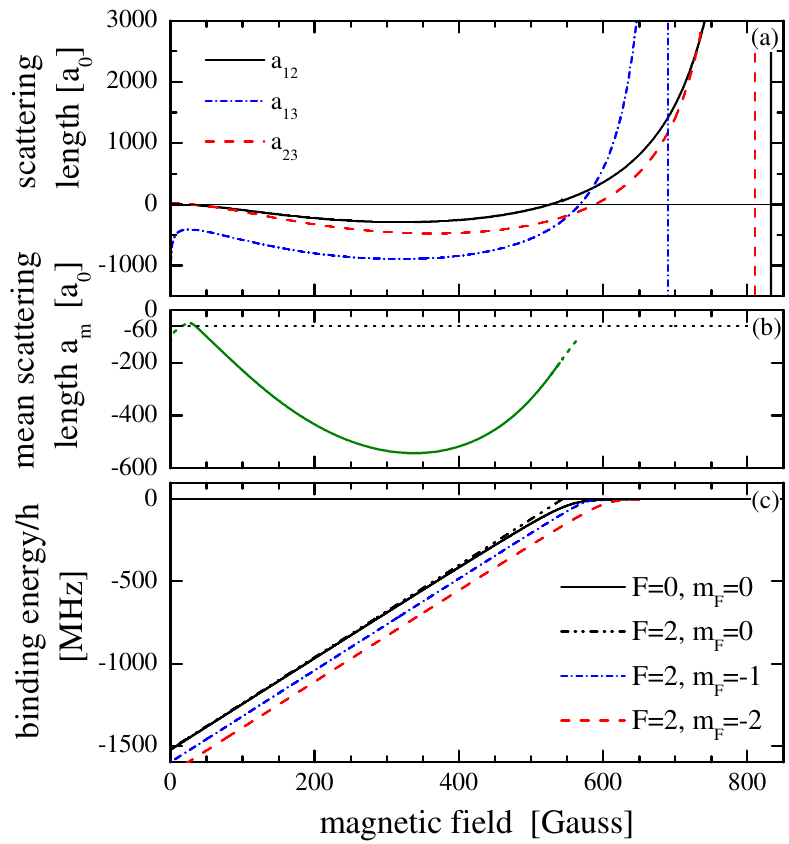}
	\caption{(Color online) (a) Magnetic field dependence of the two-body s-wave scattering lengths for different channels in units of Bohr's radius $a_0$ \cite{julienne}. (b) The solid line depicts the mean scattering length (dashed line, if one or more of the $a_{ij}>0$). For comparison, the range of the interactions ($r_0 \approx 60 a_0$) is also displayed as a dotted line. (c) Binding energies of the bound dimer states responsible for the Feshbach resonances at $543$~G, $691$\,G, $811$\,G and $834$\,G \cite{binding_energies}.}
	\label{fig:amean}
\end{figure}

In a system of three distinguishable fermions there are three different scattering lengths between the respective particles ($a_{12}$, $a_{13}$ and $a_{23}$). Yet, it is possible to use equation \ref{K_3_braaten} by combining all three scattering lengths to one single interaction parameter.
To do so, we assume that a three-body collision can be described by two independent two-body events which are described by the two-body scattering lengths (see figure \ref{fig:mediated}). 
Therefore, the probability of a three-body event is proportional to the product of the cross sections of the two-body collisions. These scale with $a_{ij}^2$ which leads to an $a_{ij}^2a_{jk}^2$-scaling for the rate of three-body collisions.
In a system of three non-identical fermions a three-body event can occur in three distinct combinations of pairwise interactions. 
By summing over these contributions we define a scaling parameter for the total rate of three-body collisions:
\begin{equation}
a_m^4 = \frac{1}{3} \left(a_{12}^2 \, a_{13}^2 + a_{12}^2 \, a_{23}^2 + a_{13}^2 \,  a_{23}^2\right).
\label{a_mean}
\end{equation}
This scaling is valid except for those cases where one scattering length diverges. 
For the three lowest hyperfine states of $^6\rm{Li}$ all scattering lengths are negative and of similar size in the magnetic field region below $550$\,G  (see figure \ref{fig:amean}(a)), thus we can define an effective mean scattering length $a_m$ as the negative solution of equation \ref{a_mean}. The value of $a_m$ as a function of the magnetic field is shown in figure \ref{fig:amean}(b). Between $100$\,G and $550$\,G $\left|a_m\right|$ is significantly larger than the range of the interactions which is given by the van der Waals length $r_0=62.5\, a_0$ for $^6$Li. Therefore, the universal theory should be applicable in this magnetic field range. 

Using our effective interaction parameter $a_m$ instead of the two-body scattering length $a$ in equation \ref{K_3_braaten}, we adapt the universal prediction for $K_3$ to the case of three non-identical fermions. For three equal scattering lengths $a_{12} = a_{13}=  a_{23} = a$ the effective interaction parameter $a_m=a$ and we recover equation \ref{K_3_braaten}.
Additionally, our ansatz correctly describes the fact that Efimov physics also occurs if only two scattering lengths are large compared to the range of the interaction $r_{0}$ \cite{braaten_review}. Then one particle can mediate the three-body interaction between the other two (see figure \ref{fig:mediated} (b)). If only one scattering length is large and the other two are small compared to $r_0$, this is not possible and thus there are no universal trimers. 
An additional justification of our ansatz for the effective interaction parameter $a_m$ are calculations using two independent models that investigate the scaling of the three-body loss coefficient with the different scattering lengths for two limiting cases of our model. Braaten $et \, al.$ calculated the three-body recombination for $a_{13}\gg a_{12},\,a_{23}$ and $a_{12} \approx a_{23}$ \cite{braaten_ourdata} while J. D'Incao and B. Esry investigated a system with $a_{13} \gg a_{23} \gg a_{12} \gg r_0$ \cite{incao, incao_overlapping}. Compared to our model these two calculations yield a similar scaling of $K_3$ with the different scattering lengths.

\begin{figure} [htbp]
\centering
	\includegraphics [width= 8cm] {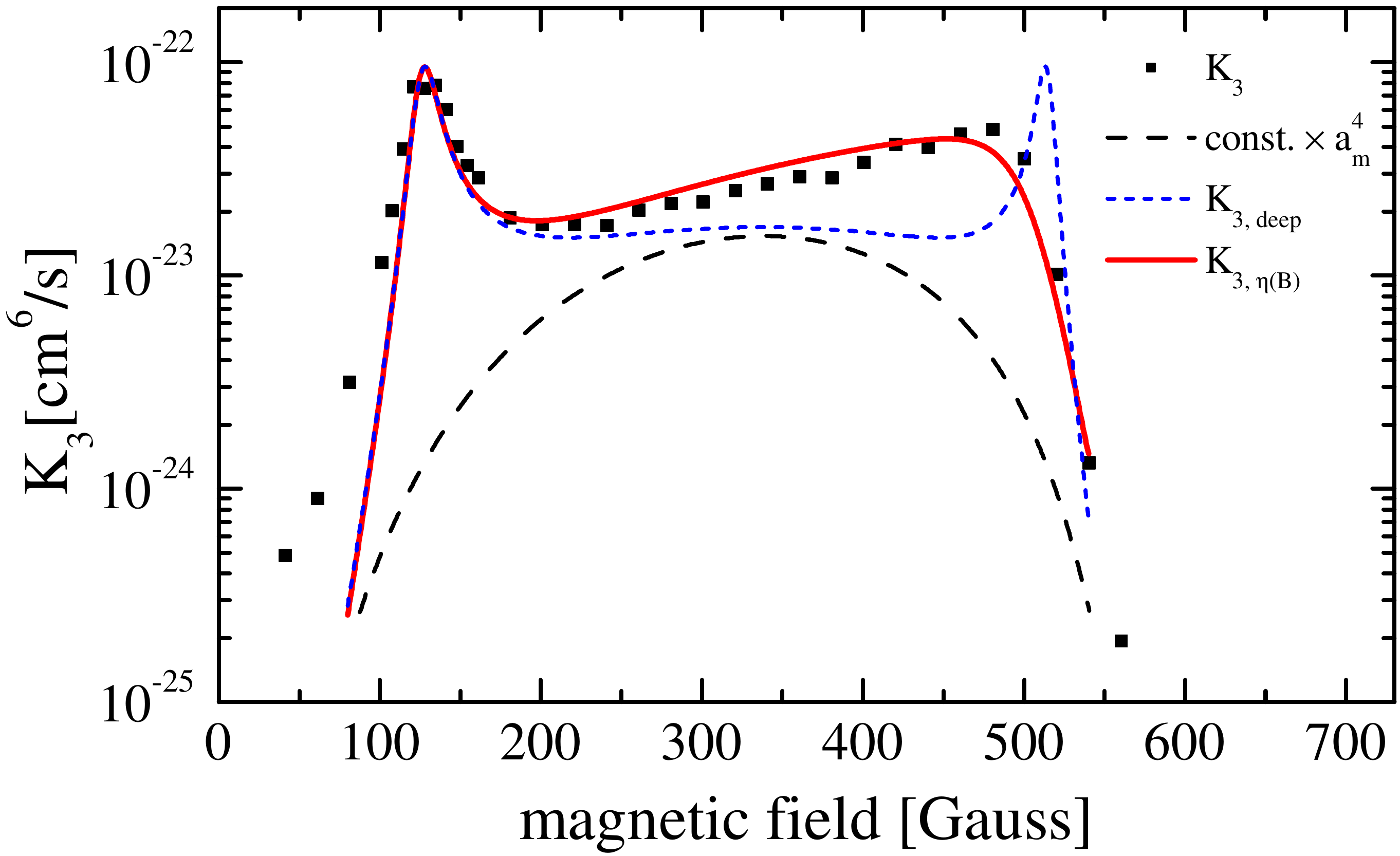} 
	\caption{(Color online) Measured three-body loss coefficient (black squares) and different models as a function of the magnetic field. The dashed line shows the prediction for three-body loss without resonant enhancement due to trimer states, the short dashed line gives the theoretical result for $K_{3,\rm{deep}} \left(a_m\right)$ and the solid line is the model for $K_{3}$ including a magnetic field dependence of $\eta_*$.}
	\label{fig:k3}
\end{figure}

If we now use our model to describe the behavior of the observed three-body loss coefficient, we get the result plotted as a short-dashed line in figure \ref{fig:k3}. The free parameters of the fit are the position of one loss resonance $a_*$, the width of this resonance $\eta_*$ and the overall amplitude $c$. The parameters were optimized to reproduce the narrow resonance at $127$\,G and the resulting values are $a_* = - 292\,a_0$, $\eta_* = 0.072$ and $c = 5612.88$. Our result is very similar to the ones obtained by several theory groups using an effective field theory model \cite{braaten_ourdata}, a hyperspherical treatment \cite{naidon_ourdata} or functional renormalization group theory \cite{wetterich_ourdata}.
However, our method did not require a rigorous treatment of all scattering lengths independently, instead we introduced an effective interaction parameter, which allows us to use an existing semi-analytical model \cite{braaten_review}. This greatly simplifies the analysis and yields very similar results as in \cite{braaten_ourdata, naidon_ourdata, wetterich_ourdata}.

Although all models with constant three-body parameters do not fully describe the experimental data, they still offer a great amount of physical insight. The good fit to the shape of the left peak suggests that the loss resonance is caused by an Efimov-like state crossing the continuum at $B=127$\,G. They also predict that this state has to become unbound again at a magnetic field of about $500$\,G, where $a_m$ is again equal to $a_*$, resulting in a second resonance caused by the same trimer state. Although the prediction of a sharp peak near $500$\,G made by these models is obviously incorrect, the comparison between the experimental data and a non-resonant behavior (long dashed line, $K_3 = \rm{const.} \times a_m^4$) indicates a strong resonant enhancement of the three-body loss measured in this magnetic field region. It was the unanimous conclusion of \cite{braaten_ourdata, naidon_ourdata, wetterich_ourdata} that a trimer state crossing the continuum at $127$\,G and close to $500$\,G is the most probable explanation for the experimental data. They also suggest that there is additional physics which changes the behavior of $K_3$ and is not considered by their models. 
This could be non-universal corrections or a magnetic field dependence of the three-body parameters. In \cite{braaten_ahoch4, braaten_deep_old}, E. Braaten and coworkers state that $a_*$ and $\eta_*$ are in principle smooth functions of the magnetic field but can be approximated by constant values near narrow Feshbach resonances. In the magnetic field range below the zero crossings (B $\lesssim 530$\,G) this condition is clearly not fulfilled and therefore a magnetic field dependence of these parameters should be considered. 

Both loss resonances at $127$\,G and near $500$\,G appear at approximately the same values of $a_m$. Thus, we neglect a dependence of the short-range three-body physics on the magnetic field which would change $a_*$. 
This is also supported by recent observations of a three-body loss resonance at approximately $900$\,G in K.~M.~O'Hara's \cite{ohara_oben} and our group \cite{wir_oben} which is most likely caused by the first excited Efimov trimer.
At this field $a_m \approx -5400\,a_0$, which is close to the value of $ 22.7 \times -292\,a_0 \approx -6630\,a_0$ expected from applying Efimov's scaling law to the three-body parameter determined by fitting the loss resonance at 127\,G. 

In contrast, the width parameter $\eta_*$ cannot be constant since the resonance near $500$\,G is significantly broader than the one near $127$\,G. The width of the resonances depends on the lifetime of the trimer, which decays into a deeply bound dimer and a free atom. Therefore, one needs to examine the magnetic field dependence of these dimer states to explain the variation of $\eta_*$ with the magnetic field. 
The dimer states closest to the continuum are the ones responsible for the Feshbach resonances shown in figure \ref{fig:amean} (a). Their binding energies can be calculated \cite{binding_energies} and are shown in figure \ref{fig:amean} (c). Away from the Feshbach resonances the states tune roughly with two Bohr magnetons compared to the free atom continuum and in the magnetic field region of interest their binding energies change by more than a factor of $5$. 
Additionally, the triplet (open-channel) admixture to the dimer states decreases with growing distance from the Feshbach resonances, which could also affect the decay probabilities.
Thus $\eta_*$, which describes the lifetime of the trimer, should vary considerably in the region of the three-body resonances. As a parameter for the change of the deeply bound dimer states we use their binding energies ($E_{B,i}$). The experimental data suggest that the lifetime of the trimer is shorter if the binding energies are smaller. 
We find that already a simple approximation $\eta_* \propto 1/E_B$ greatly improves the description of the experimental data.  
Since there are several possible decay channels due to the different dimer states, we have to sum up all different contributions. Assuming all channels contribute equally, we obtain
\begin{equation}
\eta_* = A \left( \frac{1}{E_{B,1}} + \frac{1}{E_{B,2}} + \frac{1}{E_{B,3}} + \frac{1}{E_{B,4}} \right),
\label{eta}
\end{equation} 
where $E_{B,i}$ are the different binding energies shown in figure \ref{fig:amean} (c) and A is a numerical constant \cite{deeper_dimers}. To test our model we set A such that it reproduces the value of $\eta_*$ obtained with the constant-lifetime model at the position of the narrow resonance and use the same values for resonance position and overall amplitude as in our previous fit. 
The resulting three-body loss coefficient is plotted in figure \ref{fig:k3} as a solid line and agrees remarkably well with the experimental data over the whole magnetic field range.
The fact that this agreement can be reached without modifying the three-body parameter $a_*$ suggests that the trimer state does not change significantly with the magnetic field. 


In conclusion, we have established a way to describe the strength of the three-body interaction between non-identical particles with an effective mean scattering length $a_m$. We have used this effective scattering length to describe the behavior of the three-body recombination observed in a system of atoms in three different spin states of $^6$Li with a semi-analytical model developed for identical bosons. Our model agrees well with the results from other theoretical models \cite{braaten_ourdata, naidon_ourdata, wetterich_ourdata}. Subsequently, we have expanded our model to include a variation of the width parameter $\eta_*$. This is necessary as the trimers primarily decay into a free atom and one of the most weakly bound deep dimers, whose properties change significantly in the magnetic field range of interest.  We find that the data can be described extremely well over the whole magnetic field range by assuming a $1 / E_b$-dependence of $\eta_*$ without changing any other parameters. Therefore, we can explain the observed behavior of $K_3$ with a trimer state which is universal over the whole magnetic field range. Its lifetime, however, depends on  the properties of non-universal deep dimers, which change with the magnetic field. Although the scaling $\eta_*  \propto 1/ E_b$ is purely empirical at the moment, it might be possible to calculate $\eta_*(E_b)$ explicitly since the properties of these dimers are well known from precision measurements close to the Feshbach resonances \cite{bartenstein_julienne}.
The strong dependence of $\eta_*$ on the deep dimer states which we have observed in the $^6\rm{Li}$ system suggests that such a variation of the three-body parameter $\eta_*$ should also be considered in other systems, as for example in $^{133}\rm{Cs}$ \cite{innsbruck_efimov,innsbruck_atomdimer}.

We thank E. Braaten, C. Chin, J. D'Incao, S. Fl\"orchinger, H.-W. Hammer and R. Schmidt for helpful discussions. We thank J. Ullrich and his group for their generous support. A.N.W. acknowledges support by the IMPRS-QD.

\bibliography{paper}

\begin{thebibliography}{30}
\bibitem[{dee({\natexlab{a}})}]{deep_dimer}
\bibinfo{note}{We use the term deep dimer for all non-universal two-body bound
  states. This means that their binding energy is $\gtrsim \hbar / \left(m \,
  r_0^2\right)$, where $r_0$ is the range of the interaction potential}.
\bibitem[{low()}]{low_energy}
\bibinfo{note}{{From here on we assume to be in the low-energy ($k \rightarrow
  0$) and zero-range ($r_0 \rightarrow 0$) limit.}}
\bibitem[{bin()}]{binding_energies}
\bibinfo{note}{{The binding energies of the molecular states were calculated
  using formulas derived in \cite{feshbach_review}.}}
\bibitem[{dee({\natexlab{b}})}]{deeper_dimers}
\bibinfo{note}{{In principle one has to include all deeply bound states, but
  since the binding energies of the next deeper bound dimers are orders of
  magnitude larger their relative changes are small. Thus they do not
  contribute to a magnetic field dependence of $\eta_*$ and we neglect them for
  simplicity.}}
\end{thebibliography}

\end{document}